\documentclass[manuscript]{aastex}

\shorttitle{Solar Flare Prediction Using SDO/HMI Vector Magnetic Field Data with a Machine-Learning Algorithm}
\shortauthors{Bobra and Couvidat}

\usepackage{natbib}
\bibpunct{(}{)}{;}{}{,}{,}

\begin{document}
\title{Solar Flare Prediction Using SDO/HMI Vector Magnetic Field Data with a Machine-Learning Algorithm}
\author{M. G. Bobra and S. Couvidat}
\affil{W.W. Hansen Experimental Physics Laboratory, Stanford University, Stanford, CA 94305}

\begin{abstract}
We attempt to forecast M-and X-class solar flares using a machine-learning algorithm, called Support Vector Machine (SVM), and four years of data from the Solar Dynamics Observatory's Helioseismic and Magnetic Imager, the first instrument to continuously map the full-disk photospheric vector magnetic field from space. Most flare forecasting efforts described in the literature use either line-of-sight magnetograms or a relatively small number of ground-based vector magnetograms. This is the first time a large dataset of vector magnetograms has been used to forecast solar flares. We build a catalog of flaring and non-flaring active regions sampled from a database of 2,071 active regions, comprised of 1.5 million active region patches of vector magnetic field data, and characterize each active region by 25 parameters. We then train and test the machine-learning algorithm and we estimate its performances using forecast verification metrics with an emphasis on the True Skill Statistic (TSS). We obtain relatively high TSS scores and overall predictive abilities. We surmise that this is partly due to fine-tuning the SVM for this purpose and also to an advantageous set of features that can only be calculated from vector magnetic field data. We also apply a feature selection algorithm to determine which of our 25 features are useful for discriminating between flaring and non-flaring active regions and conclude that only a handful are needed for good predictive abilities.
\pagebreak
\end{abstract}

\section{Introduction}
Though the energy release mechanisms of solar flares have yet to be fully characterized, they are certainly magnetic in nature \citep{ForbesPriest2002}. Thus, studying the magnetic field configuration in the solar atmosphere is critical to understanding and ultimately predicting solar flares. Even though the magnetic field in the corona cannot be mapped directly, the magnetic field at the photosphere can. Until recently, most photospheric magnetic field data contained only the line-of-sight component of the magnetic field; the full vector had been mapped non-continuously or for only part of the solar disk. The Solar Dynamics Observatory's Helioseismic and Magnetic Imager (HMI) is the first instrument to continuously map the full-disk photospheric vector magnetic field \citep{schou12}. Since May 2010, HMI has mapped the vector magnetic field every 12 minutes 98.44\% of the time \citep{hoeksema14}. 

Many flare prediction studies involve using photospheric magnetic field data to parameterize active regions such that they can be described by a few numbers. The active region (AR) parameters considered in the literature are varied; some characterize the magnetic field topology \citep[e.g.][]{schrijver07}, others measure the integrated Lorentz force exerted by an AR (e.g. Fisher et al., 2012), and still others parameterize energy, helicity, currents, and shear angles \citep[e.g.][]{moore12,labonte07,lb03,hagyard84}. All of these parameters were developed with the ultimate goal of finding some relationship between the behavior of the photospheric magnetic field and solar activity, which typically occurs in the chromosphere and transition region of the solar corona. 

However, the relationship between the photospheric and coronal magnetic field during a solar flare is not fully understood. As such, flare prediction has been so far mostly based on the use of classifiers trying to automatically find such a relationship rather than on purely theoretical considerations from which such a relationship could be derived. Classifiers are algorithms that, with proper training, predict which class an example most likely belongs to. In the context of solar flare predictions, we may restrict the classes to two, \mbox{i.e.} a binary classifier. For example, an active region belongs to the positive class if it produces one or more flares within a given time interval. Conversely, an active region belongs to the negative class if it does not produce a flare in the same time interval. Classifiers can be either linear or non-linear: in the feature-space, \mbox{i.e.} the space of the parameters characterizing each example of flaring and non-flaring active regions, a linear classifier tries to find an hyperplane that best separates both classes. For a linear classifier to perform well, the two classes have to be linearly separable. However, it is not clear that flaring and non-flaring ARs can be linearly separated for a given set of features. 

Also, there are several machine learning algorithms specifically designed as non-linear classifiers. Machine learning is a branch of computer science intent on developing algorithms that can automatically learn from the input data. Supervised machine-learning classifiers are provided with a set of examples and their corresponding class: this set is used for training the classifier through a learning algorithm. 

\citet{li07}, \citet{colak07}, \citet{song09}, \citet{yu09}, \citet{yuan10} and \citet{ahmed} used non-linear machine-learning algorithms to forecast solar flares. These studies used light-of-sight magnetic field data, solar radio flux, or metadata (\mbox{e.g.} McIntosh class, sunspot number) to characterize their features. \citet{lb03} pioneered the use of vector magnetic field data for flare prediction. They used the largest database of vector magnetic field data, from the Mees Solar Observatory Imaging Vector Magnetograph on the summit of Mount Haleakala, available at the time with a prediction technique known as discriminant analysis, which is a linear classifier. Other prediction techniques include superposed epoch analysis \citep{mason10} and statistical analyses \citep[e.g.][]{bloomfield12,falconer12}. 

A continuous stream of vector magnetograms provide more information about the photospheric magnetic field topology than line-of-sight data and thus is expected to lead to better predictive capabilities. In this paper, we systematically use HMI vector magnetograms with a non-linear classification algorithm, called Support Vector Machine, to attempt to predict solar flares. This is the first time such a large dataset of vector magnetograms has been used to forecast solar flares. 

The paper is organized as follows: In Section \ref{section:catalog}, we describe our flare catalog and active region parameters. In Section \ref{section:ml}, we describe the machine learning algorithm. In Section \ref{section:metrics}, we discuss metrics to measure a classifier's performance, particularly in a class-imbalanced problem. And finally, in Section \ref{section:results}, we present our results and compare our metric scores with those who conducted similar analyses.

\section{Flare selection}
\label{section:catalog}

To train a flare forecasting algorithm, we need a catalog of flaring and non-flaring ARs, respectively called positive and negative examples. To build this dataset, we only consider flares with a Geostationary Operational Environmental Satellite (GOES) X-ray flux peak magnitude above the M1.0 level, \mbox{i.e.} only major flares. We reject C-class flares because a significant number of C-class flares in the GOES data are not associated with NOAA active-region numbers, which makes it difficult to pinpoint their location. It is not clear what the impact of including or rejecting C-class flares from the catalog is on the performance of the forecasting algorithm. \citet{bloomfield12} highlight in their Table 4 that including C-class flares may improve some performance metrics while lowering others: in their case, not including C-class flares increase the True Skill Statistic (TSS) metric but decrease the Heidke Skill Score (HSS) (both quantities are described in Section \ref{section:metrics}).

\subsection{GOES data}

We use the peak magnitude of the X-ray flux observed by GOES to classify
solar flares. We only consider flares (i) with a peak magnitude of
10$^{-5}$ Watts m$^{-2}$, which correspond to M1.0-class or higher, (ii)
that occur within $\pm$ 68$^{\circ}$ of the central meridian, and (iii)
whose locations are identified in the National Geophysical Data Center
GOES X-ray flux flare catalogs. As shown in Figure 6 of
\citet{bobra}, the signal-to-noise in the SHARP parameters, described in
Section \ref{subsection:sharpmask}, increases significantly beyond $\pm$
70$^{\circ}$ of central meridian. As such, we only considered flares
that occur within $\pm$ 70$^{\circ}$ of central meridian. Upon imposing
this restriction, we found that the furthest X-class flare in our sample
occurred at 68$^{\circ}$ from central meridian. Thus, we cut off our
analysis at this longitude.  It is also important to note that we do
consider every flare produced by any given active region; thus, if an
active region flared 10 times, we count that as 10 separate flares. As
such, our sample contains 285 M-class flares and 18 X-class flares
observed between May 2010 and May 2014.

\subsection{Definitions of positive and negative classes}
\label{subsection:classes}

In order to train our classifier, we must clearly define the positive and negative classes. In this study, we follow the definition outlined in \citet{ahmed}. They posited two forms of associations between active regions and flares --- operational and segmented forms. For the operational case, an active region that flares within 24 hours after a sample time belongs to the positive class. Conversely, an active region that does not flare within 24 hours after a sample time belongs in the negative class. The positive class is defined in the same way for the segmented case; however, the negative class is defined differently: if an active region does not flare within a $\pm$ 48 hour period from the sample time, it belongs to the negative class. 

For flaring active regions, the sample time is defined to be exactly 24 hours prior to the GOES X-ray flux peak time by construction. For flare-quiet times, the sample time is chosen randomly. Our catalog includes the 303 positive examples previously mentioned (285 M-class flares and 18 X-class ones), and 5000 randomly selected negative examples. The number of flaring ARs is relatively small, especially compared to previous studies based on data from the Solar and Heliospheric Observatory's Michelson Doppler Imager \citep[e.g.][]{ahmed}. This is the result of the unusually quiet cycle 24 that is contemporaneous with HMI observations.

\subsection{Active region features: the SHARP parameters} 
\label{subsection:sharpmask} 

The HMI data repository, located at the Joint Science Operations Center\footnote{see http://jsoc.stanford.edu}, contains the first continuous measurement of the full-disk photospheric vector magnetic field taken from space \citep{schou12}. In 2014, the HMI team released a derivative data product, called Space-weather HMI Active Region Patches (SHARP), which automatically identifies AR patches in the vector magnetic field data \citep{bobra}. These patches are then tracked as they cross the solar disk. The SHARP data set provides a number of benefits, one of which is the ability to automatically calculate AR summary parameters on a twelve-minute cadence continuously throughout the HMI mission. 

\begin{figure}
\renewcommand{\tabcolsep}{0.0018\textwidth}
\begin{tabular}{ccc}
\includegraphics[angle=90,width=0.496\textwidth]{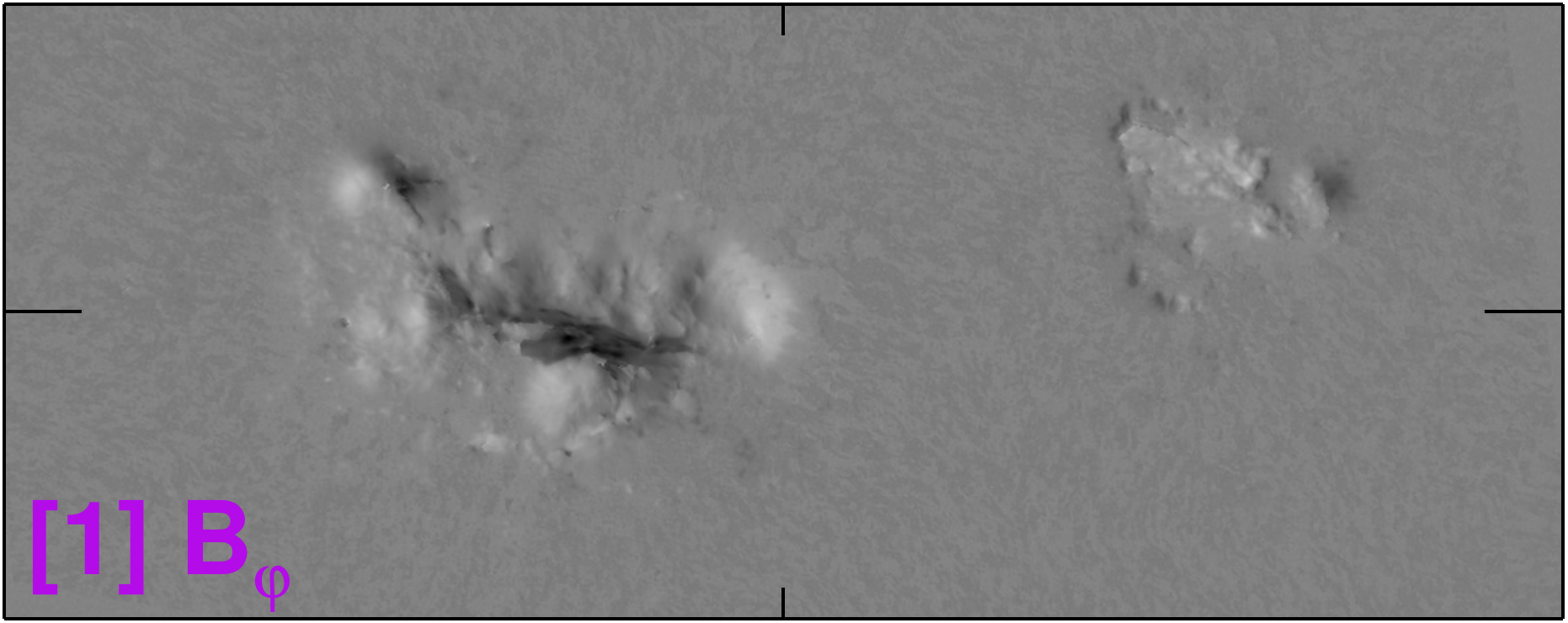} &
\includegraphics[angle=90,width=0.496\textwidth]{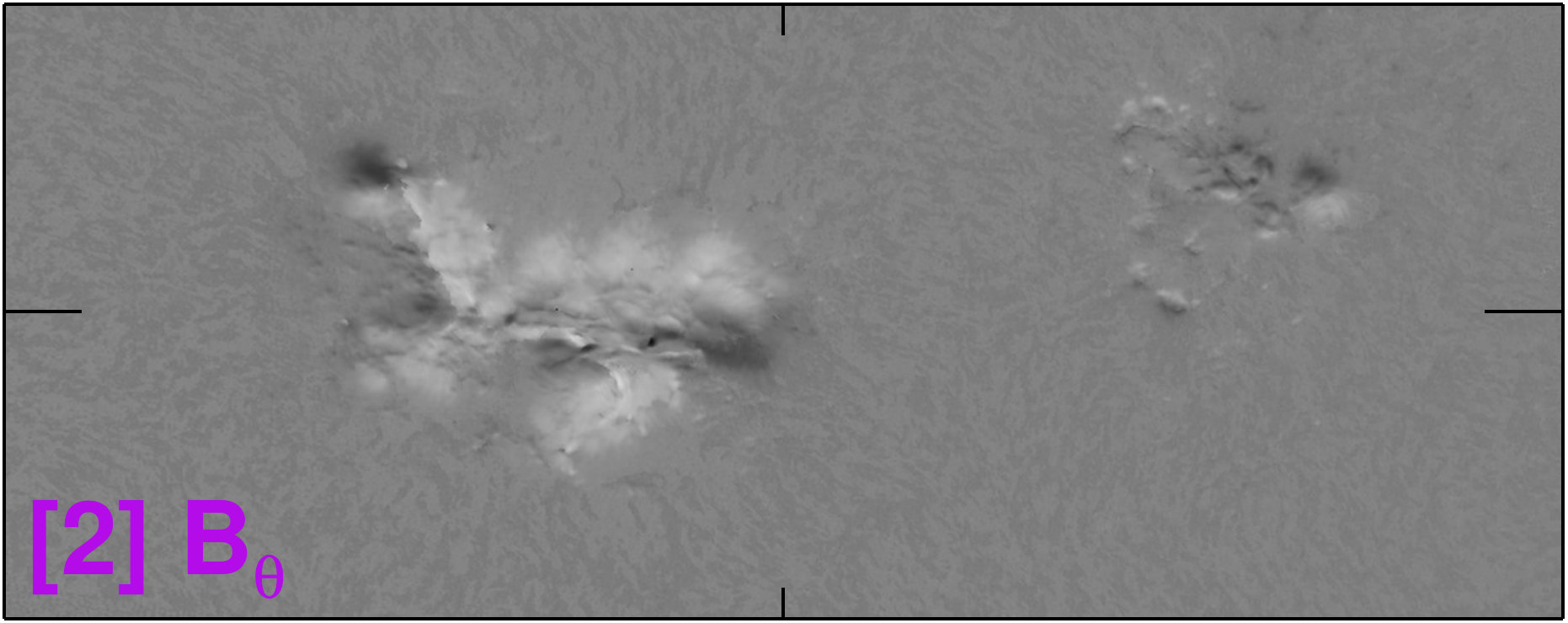} \\
\includegraphics[angle=90,width=0.496\textwidth]{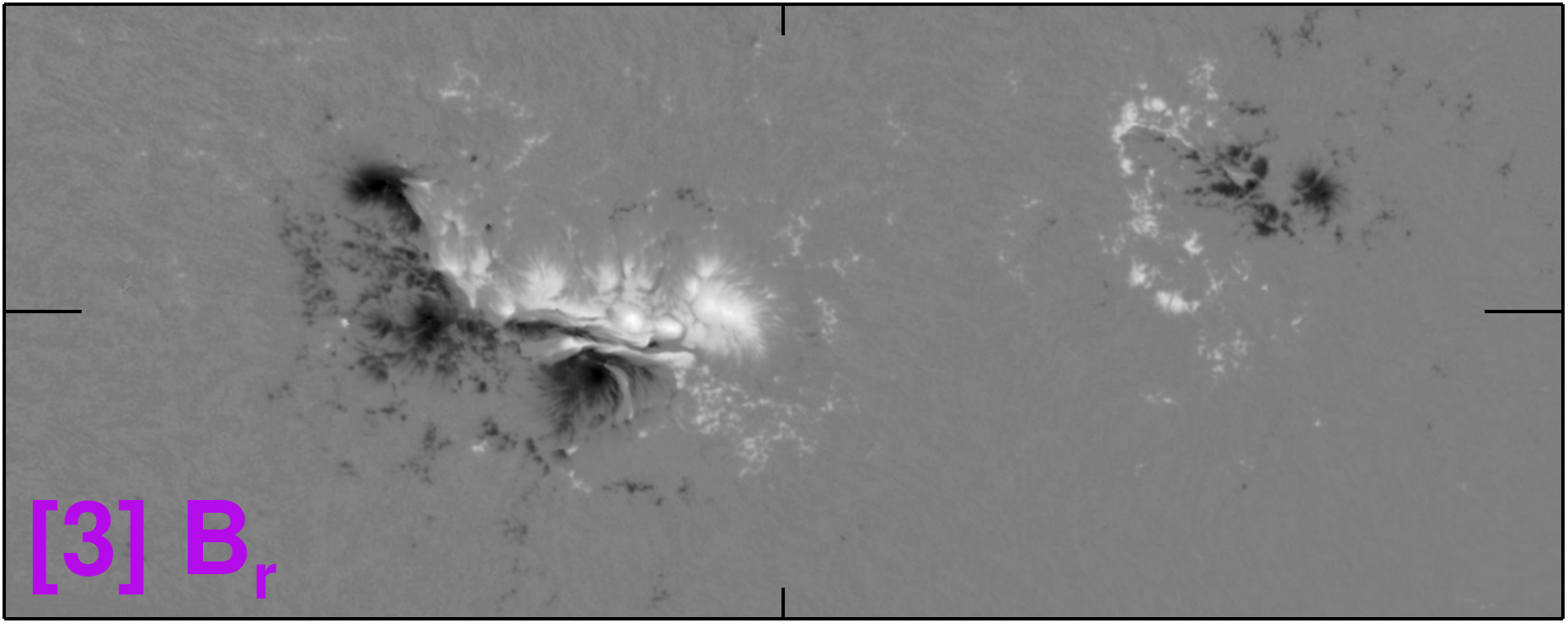} &
\includegraphics[angle=90,width=0.496\textwidth]{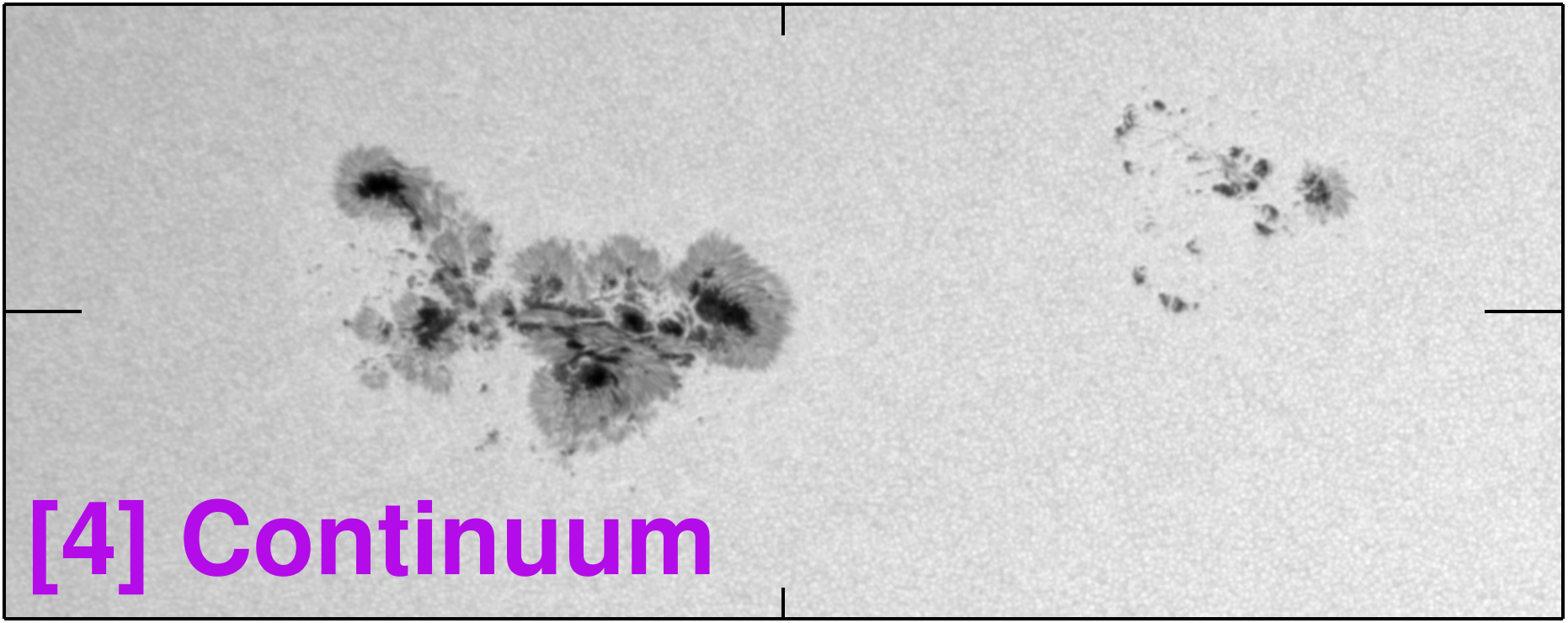} \\
\includegraphics[angle=90,width=0.496\textwidth]{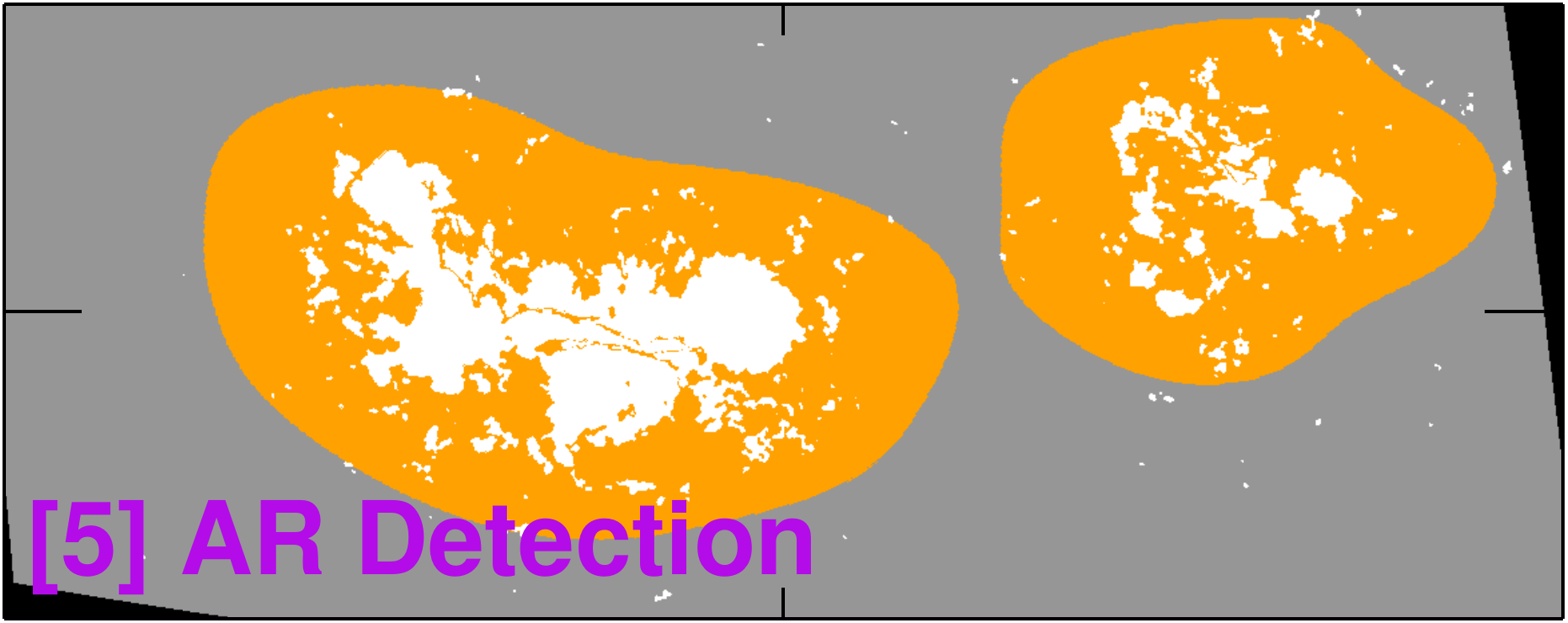} &
\includegraphics[angle=90,width=0.496\textwidth]{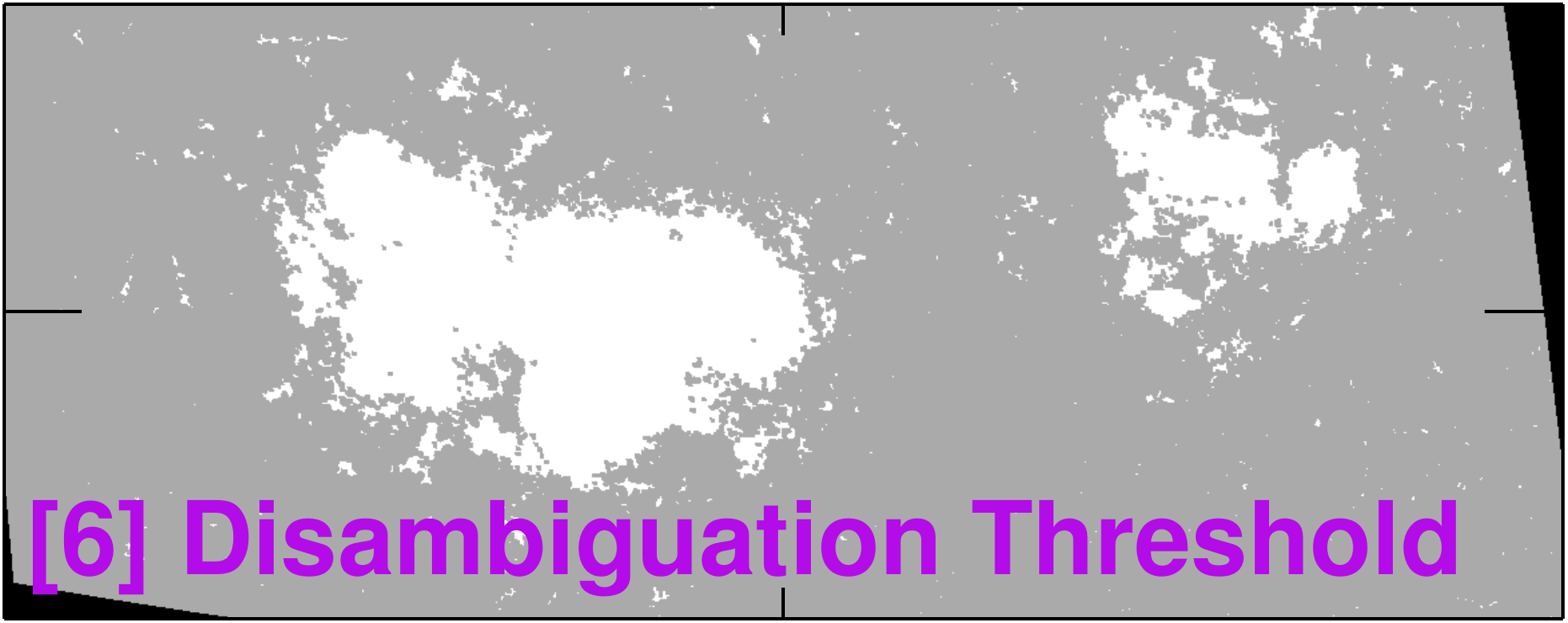} \\
\includegraphics[angle=90,width=0.496\textwidth]{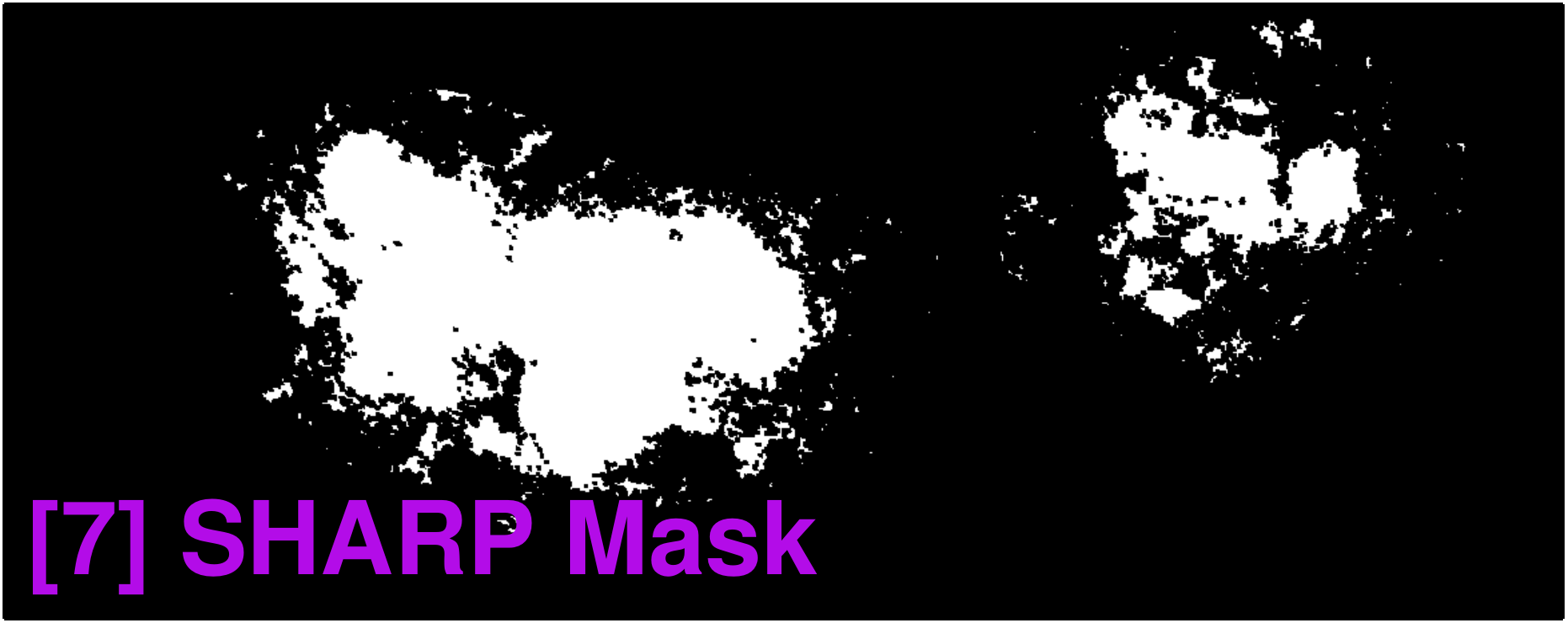} &
\includegraphics[angle=90,width=0.496\textwidth]{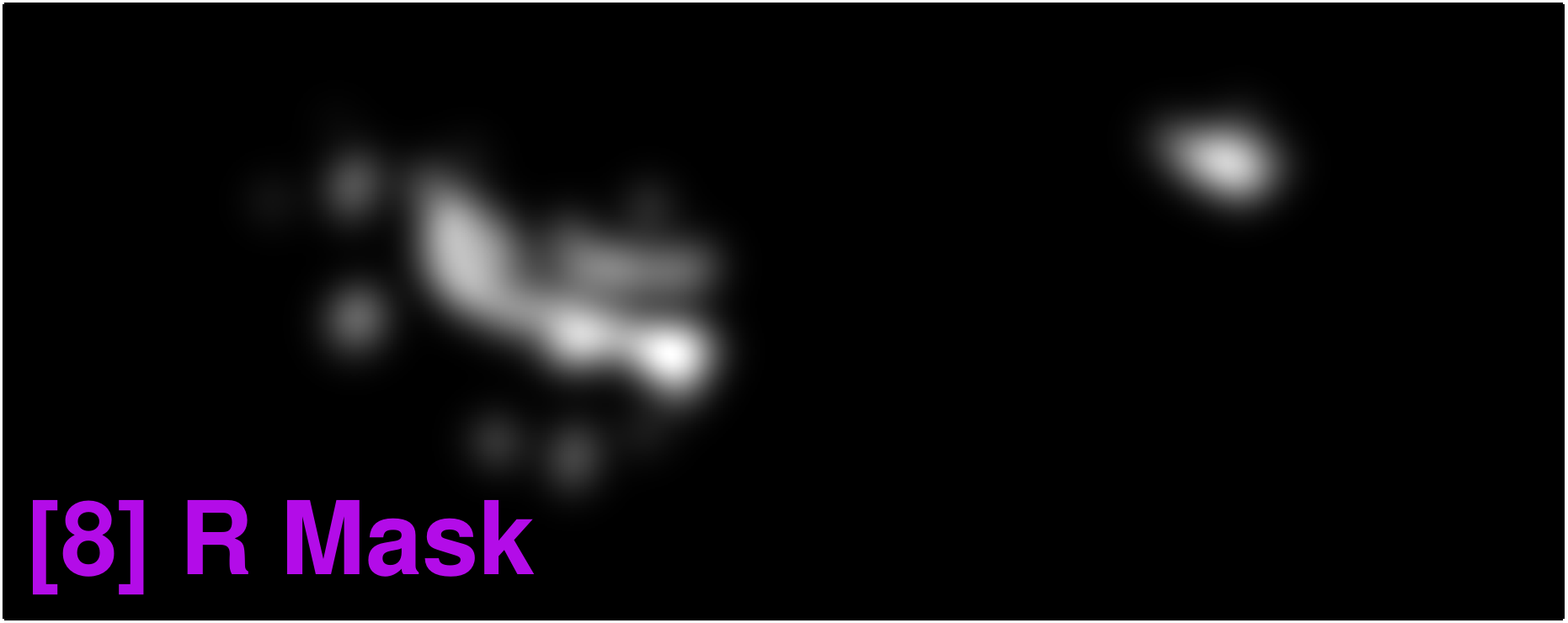} \\
\end{tabular}
\caption{The first four panels show each of the components of the vector magnetic field data, $B_{\phi}$, $B_{\theta}$, and $B_r$, and the continuum intensity data for NOAA Active Region 11429 on March 7, 2012 at 00:24 TAI. The color table is scaled between $\pm$ -2500 Gauss for all three magnetic field arrays. The fifth panel shows the result of the active region automatic detection algorithm; the sixth panel shows pixels above the high-confidence disambiguation threshold. Only pixels that are within the orange-shaded region and above the high-confidence disambiguation threshold contribute to the calculation of the active region SHARP parameters. These pixels are shown in the seventh panel. The eighth panel shows the result of the polarity inversion line automatic detection algorithm. These pixels contribute to the active region parameter R. All the panels are in the Cylindrical Equal-Area coordinate system and centered on CEA Longitude $304.7^\circ$ and CEA Latitude $17.5^\circ$ in Carrington Rotation 2120; the patches span $35^\circ$ in CEA Longitude and $14^\circ$ in CEA Latitude.}
\label{fig:example}
\end{figure}

For this study, we calculate 25 parameters, or features, using four years of SHARP vector magnetic field data, from May 2010 to May 2014. Each parameter is calculated every 12 minutes during an AR lifetime. Within this four-year time period, the SHARP database contains 2,071 ARs and 1.5 million active region patches of vector magnetic field data, comprising a total of 37.5 million unique parameterizations of the photospheric vector magnetic field. The 25 features characterizing any given active region are listed in Table \ref{tbl-3}, along with a brief description and formula. We parameterize various physical quantities, such as the total flux through the surface area of the active region. We also calculate proxies for other physical quantities, such as helicity and energy. It is worth noting that the free energy proxy is sensitive to the choice of potential field model; we use a fast Fourier transform method. We also calculate various geometric quantities, such as the inclination angle, and topological quantities, such as the flux within the polarity inversion line. Each parameter is calculated using one of two pixel masks, shown in panels 7 and 8 of Figure \ref{fig:example}.  Only the calculations for R\_VALUE and AREA\_ACR use the line-of-sight magnetic field; the rest of the parameters in this study are calculated using components of the vector magnetic field. 

The SHARP parameters are adapted from myriad studies. Sixteen of these parameters come from \citet{lb03}. Seven of these parameters characterize the Lorentz force exerted by an AR on the solar atmosphere and are adapted from \citet{fisher12}, who find that solar activity may be associated with a Lorentz force impulse at the photosphere. We employ an algorithm that uses a Bayesian inversion method to automatically detect ARs in line-of-sight magnetic field data \citep{turmon10}. The AREA\_ACR parameter, which characterizes the de-projected area of the strong-field pixels, is an output of this automatic AR detection algorithm. The R\_VALUE parameter, named by \citet{schrijver07} as simply R, characterizes the total unsigned flux near high-gradient AR polarity inversion lines. We calculate R using the exact same methodology as \citet{schrijver07}, who used the line-of-sight component of the magnetic field.

Figure \ref{fig:example} shows SHARP data for NOAA Active Region 1449, which produced an X5.4-class flare on March 7, 2012 at 00:24:00 TAI. The first three panels show the inverted and disambiguated data wherein the vector {\bf B} has been remapped to a Cylindrical Equal-Area (CEA) projection and decomposed into  $B_{\phi}$, $B_{\theta}$, and $B_r$, respectively, in standard heliographic spherical coordinates [$\bf{\hat{e}_r}$, $\bf{\hat{e}_\theta}$, $\bf{\hat{e}_\phi}$] following Equation 1 of \citet{gary90}. The fourth panel shows the continuum intensity data for the same region at the same time. The fifth panel shows the result of the AR automatic detection algorithm employed to create the SHARP data series \citep{turmon10}. This detection algorithm operates on the line-of-sight magnetic field images and creates a bitmap to encode membership in the orange-colored coherent magnetic structure. As such, the detection algorithm's definition of an AR is not necessarily the same as NOAA's definition of an AR. At times, the detection algorithm will combine into one AR what NOAA defines as multiple ARs. 

All of the AR parameters, except for R, are calculated on the pixels identified as white in panel seven of Figure \ref{fig:example}. These pixels are defined as both those that (i) reside within the orange-colored magnetic structure identified in panel five and (ii) satisfy a high-confidence disambiguation threshold (indicated by the white pixels in panel six; see \citet{hoeksema14} for more details). R is calculated using the result of an algorithm that is designed to automatically identify the polarity inversion line \citep{schrijver07}, the result of which is shown in the eighth panel of Figure \ref{fig:example}. 

It is important to note that the AR parameters are highly sensitive to which pixels contribute to their calculation. The mask represented by panel seven may not be the best choice due to the presence of some weak-field pixels, which contain low signal-to-noise. It is worthwhile to study how to optimize this mask such that it yields the strongest pre-flare signature per AR parameter. It is also noticeable that the AR parameters are sensitive to the periodicity in magnetic field strength due to the orbital velocity of SDO. This periodicity is described in detail in Section 7.1.2 of \citet{hoeksema14}.  Finally, the parameters are slightly sensitive to the errors introduced by mapping the vector magnetic field data from CCD to CEA coordinates, which can be estimated using Equation 9 in \citet{sun13}. In general, the deviation between the true vector {\bf B} and the mapped vector {\bf B} is less than a few degrees for ARs that are less than 45 degrees square and near central meridian. As previously mentioned, we reject all ARs outside of the area within $\pm$ 68$^{\circ}$ of the central meridian.

\begin{deluxetable}{lllll}
\tabletypesize{\scriptsize}
\tablecaption{SHARP active region parameter formulae. \label{tbl-3}}
\tablewidth{0pt}
\tablehead{\colhead{Keyword} & \colhead{Description} &  \colhead{Formula} & \colhead{F-Score} & \colhead{Selection}}
\startdata
{\sc totusjh} & Total unsigned current helicity  & ${H_{c_{total}}} \propto \sum |B_z \cdot J_z|$ & 3560 & Included \\
{\sc totbsq}  & Total magnitude of Lorentz force & $F  \propto \sum B^{2} $ &3051 & Included  \\
{\sc totpot} & Total photospheric magnetic free energy density & $ \rho_{tot} \propto  \sum \left( \vec{\textit{\textbf B}}^{\rm Obs} - \vec{\textit{\textbf B}}^{\rm Pot} \right)^2 dA $ & 2996 &Included\\
{\sc totusjz} & Total unsigned vertical current & ${J_{z_{total}}} =  \sum |J_{z}|dA$  & 2733 &Included\\
{\sc absnjzh} & Absolute value of the net current helicity & ${H_{c_{abs}}} \propto \left| \sum B_z \cdot J_z \right|$ & 2618 & Included \\
{\sc savncpp} & Sum of the modulus of the net current per polarity & $J_{z_{sum}} \propto \Big\vert \displaystyle\sum\limits^{B{_z^+}} J{_z}dA \Big\vert + \Big\vert \displaystyle\sum\limits^{B{_z^-}} J{_z}dA \Big\vert $ & 2448 & Included\\
{\sc usflux} & Total unsigned flux & $\Phi = \sum|B_{z}|dA$ & 2437 & Included \\
{\sc area\_acr} &  Area of strong field pixels in the active region &  Area $ = \sum$ Pixels &2047&Included\\
{\sc totfz}  &  Sum of z-component of Lorentz force & $F_{z}  \propto \sum (B_{x}^{2} + B_{y}^{2} - B_{z}^{2}) dA$ & 1371 &Included \\
{\sc meanpot} & Mean photospheric magnetic free energy & $ \overline{\rho} \propto \frac{1}{N} \sum \left( \vec{\textit{\textbf B}}^{\rm Obs} - \vec{\textit{\textbf B}}^{\rm Pot} \right)^2 $ & 1064 &Included\\
{\sc r\_value} &  Sum of flux near polarity inversion line &  $\Phi = \sum|B_{LoS}|dA$ within R mask &1057&Included\\
{\sc epsz}  & Sum of z-component of normalized Lorentz force  & $\delta F_{z}  \propto \frac{\sum (B_{x}^{2} + B_{y}^{2} - B_{z}^{2})}{ \sum B^{2}}$  &864.1 &  Included\\
{\sc shrgt45} & Fraction of Area with Shear $> 45^\circ$  & Area with Shear $>45^\circ$ / Total Area & 740.8 & Included\\
{\sc meanshr} & Mean shear angle  & $ \overline{\Gamma} = \frac{1}{N} \sum \arccos \left( \frac{\vec{\textit{\textbf B}}^{\rm Obs} \cdot \vec{\textit{\textbf B}}^{\rm Pot}}{|B^{\rm Obs}|\,|B^{\rm Pot}|} \right)$  & 727.9 & Discarded\\
{\sc meangam} & Mean angle of field from radial &  $\overline{\gamma} = \frac{1}{N} \sum \arctan\left(\frac{B_h}{B_z}\right)$  & 573.3 & Discarded\\
{\sc meangbt} & Mean gradient of total field & $\overline{\left|{\nabla B_{\rm tot}}\right|} = \frac{1}{N} \sum \sqrt{\left(\frac{\partial B}{\partial x}\right)^2 + \left(\frac{\partial B}{\partial y}\right)^2}$ & 192.3 & Discarded\\
{\sc meangbz} & Mean gradient of vertical field & $\overline{\left|{\nabla B_z}\right|} = \frac{1}{N} \sum \sqrt{\left(\frac{\partial B_z}{\partial x}\right)^2 + \left(\frac{\partial B_z}{\partial y}\right)^2}$ & 88.40 & Discarded \\
{\sc meangbh} & Mean gradient of horizontal field & $\overline{\left|{\nabla B_h}\right|} = \frac{1}{N} \sum \sqrt{\left(\frac{\partial B_h}{\partial x}\right)^2 + \left(\frac{\partial B_h}{\partial y}\right)^2}$ & 79.40 & Discarded\\
{\sc meanjzh} & Mean current helicity ($B_{z}$ contribution) & $\overline{H_c} \propto \frac{1}{N} \sum B_z \cdot J_z $ & 46.73 & Discarded\\
{\sc totfy}  &  Sum of y-component of Lorentz force & $F_{y} \propto \sum B_{y} B_{z} dA$  & 28.92 & Discarded\\
{\sc meanjzd} & Mean vertical current density & $\overline{J_z} \propto \frac{1}{N} \sum \left(\frac{\partial B_y}{\partial x} - \frac{\partial B_x}{\partial y}\right) $ & 17.44 & Discarded\\
{\sc meanalp} & Mean characteristic twist parameter, $\alpha$ & $ \alpha_{total} \propto \frac{\sum J{_z} \cdot B_z}{\sum B{^2_z}} $ & 10.41 & Discarded \\
{\sc totfx} &  Sum of x-component of Lorentz force & $F_{x} \propto - \sum B_{x} B_{z} dA$ & 6.147 & Discarded \\
{\sc epsy}  & Sum of y-component of normalized Lorentz force & $\delta F_{y}  \propto \frac{-\sum B_{y} B_{z}}{ \sum B^{2}}$ & 0.647 & Discarded \\
{\sc epsx}  & Sum of x-component of normalized Lorentz force & $\delta F_{x}  \propto \frac{\sum B_{x} B_{z}}{ \sum B^{2}}$  & 0.366 &Discarded \\
\enddata
\end{deluxetable}


\section{Machine-learning algorithms}
\label{section:ml}

To automatically predict whether an AR will flare or not, we apply machine learning (ML) classifiers to the SHARP datasets of positive and negative examples. Machine learning is a field of computer science that develops algorithms with the ability to learn a specific task without being explicitly programmed for it. It is commonly used for classification, regression, and clustering tasks. ML classifiers need to be trained on a catalog of examples, and tested on another catalog.

\subsection{Feature selection algorithm}

Each example of a flaring or non-flaring AR is characterized by a feature vector of SHARP parameters. High dimensionality in this vector may result in lower performance for a classifier. It is beneficial to reduce the dimensionality by getting rid of features that are not very helpful at the classification task. Here, we use a univariate feature selection algorithm with an F-score for feature scoring. This kind of feature selection algorithm is a so-called filter method \citep[e.g.][]{guyon03}, because it is applied prior to running, and is independent of the classifier (in contrast to, \mbox{e.g.}, wrapper or embedded methods). Filter algorithms involve the computation of a relevance score. They are widely used and computationally efficient. Univariate feature selection algorithms assume that the features are independent, and ignore any correlation between them. The F-score, or Fisher ranking score, $F(i)$, for feature $i$ is defined as \citep[e.g.][]{gu12}:

\begin{equation}
F(i)=\frac{(\bar{x}^+_i - \bar{x}_i)^2+(\bar{x}^-_i - \bar{x}_i)^2}{\frac{1}{n^+-1}\sum_{k=1}^{n^+}(x^+_{k,i} - \bar{x}_i)^2 + \frac{1}{n^--1}\sum_{k=1}^{n^-}(x^-_{k,i} - \bar{x}_i)^2}
\end{equation}
where $\bar{x}^+_i$ is the average of the values of feature $i$ over the positive-class examples, $\bar{x}^-_i$ is the average of the values over the negative-class examples, $\bar{x}_i$ is the average of the values over the entire dataset, $n^+$ is the total number of positives examples in the dataset, $n^-$ is the number of negative examples, and the denominator is the sum of the variances of the values of feature $i$ over the positive and negative examples separately. The F-score measures the distance between the two classes for a given feature (inter-class distance), divided by the sum of the variances for this feature (intra-class distance). We compute the F-score of each feature, and only select those with the highest score.

Here we compute $F(i)$ through the SelectKBest class of the Scikit-Learn module for the Python programming language \citep{pedregosa11}. Scikit-Learn uses the Libsvm library internally.

\subsection{The classifier: a Support Vector Machine}

We focus on the Support Vector Machine (SVM) \citep{cortes95}, which is a binary classifier. Typically, the two classes are called positive and negative. By convention, and as mentioned in Section \ref{subsection:classes}, we assign the positive class to flaring active regions, and the negative class to non-flaring active regions. To train and test the SVM, we randomly separate the catalog of positive and negative examples into two, non-overlapping, datasets.

\citet{li07} and \citet{yuan10} used a soft margin SVM algorithm to forecast solar flares, demonstrating the feasibility of this approach. Here, we use the Scikit-Learn module implementation of a soft margin SVM in the Python programming language. This implementation is both robust and fast. The SVM can be either linear or non-linear. A linear classifier seeks to separate the examples by finding an hyperplane in the feature space. To use non-linear decision functions to separate the examples, \mbox{i.e.} not an hyperplane but a more complicated hyper-surface, we can use kernels that remap the feature space into a higher-dimensional space. Whether in the original feature space or in the remapped one, the SVM tries to find the separating hyperplane with  the largest distance to the nearest training examples: it is a maximum margin classifier. A SVM solves the following quadratic optimization problem:

\begin{equation}
\mathrm{min} \biggl( \frac{1}{2} \mathbf{\omega}^T \mathbf{\omega} + C \sum_{k=1}^{m} \epsilon_k \biggr) \label{eq1}
\end{equation}
where we minimize for the vectors $\bf{\omega}$ and $\bf{\epsilon}$ subject to:
\begin{equation}
y_k(\mathbf{\omega}^T \phi(\mathbf{x_k}) + b) \ge 1-\epsilon_k
\end{equation}
and
\begin{equation}
\epsilon_k \ge 0
\end{equation}
for each training example $k$.
$m$ is the number of training examples, $C$ is a penalty parameter (soft margin parameter), $\bf{\omega}$ is the normal vector to the separating hyperplane, $\epsilon_k$ is the (slack) variable measuring the degree of misclassification of training example $k$, $y_k$ is the label of example $k$ (\mbox{i.e.} the class to which the example belongs) which can be either $+1$ (positive example) or $-1$ (negative one), $\bf{x_k}$ is the feature vector of the training example, $^T$ denotes the transpose, and $\phi$ is a function mapping the input data into a higher-dimensional space.
The kernel $k(\mathbf{x_k},\mathbf{x_k^{\prime}})$ is defined as $k(\mathbf{x_k},\mathbf{x_k^{\prime}}) = \phi(\mathbf{x_k})^T \phi(\mathbf{x_k^{\prime}})$.
Here, we use a radial basis function (or Gaussian kernel), expressed as:
\begin{equation}
k(\mathbf{x_k},\mathbf{x_k^{\prime}}) = \mathrm{exp}(-\gamma ||\mathbf{x_k}-\mathbf{x_k^{\prime}}||^2)
\end{equation}
where $\gamma$ represents the width of the kernel.

Therefore, two parameters have to be set to train the SVM: $C$ and $\gamma$.
The optimization is a trade-off between a large margin (small $||\bf{\omega}||$) and a small error penalty (small $||\bf{\epsilon}||$). The SVM is trained until the cost function of Equation \ref{eq1} varies by less than a specified tolerance level.

It is worth mentioning that we initially implemented a feed-forward neural network with a stochastic backpropagation training rule \citep[e.g.][]{lecun98}. However, its performance proved inferior to that of a SVM algorithm, as has been confirmed by numerous studies in various fields of research \citep[e.g.][]{mog01}.

\section{Metrics to Measure a Classifier's Performance in a Class-Imbalanced Problem}
\label{section:metrics}

It is essential to be able to measure the performance of a forecasting algorithm. This is no small task as the literature provides many different metrics, each one with advantages and drawbacks, including those that do not necessarily apply to the problem at hand.

\subsection{Class imbalance problem}

In the solar-flare forecasting field, the two classes (non-flaring and flaring ARs) are strongly imbalanced: there are many more negative examples than positive ones, which reflects the fact that most ARs do not produce major flares in any given 24 or 48-hour period. This class imbalance is a major issue for most machine-learning algorithms. Indeed, a ML classifier may strongly favor the majority class, and neglect the minority one. In other words, always predicting that an AR will not flare is likely to give very good results overall. Several solutions are presented in the literature to remedy this issue \citep[e.g.][]{long13}. The easiest way is to undersample the majority class: in other words, to build a training set that has about as many negative as positive examples. However, for flare prediction this gives poor results, as non-flaring ARs can exhibit quite a large range of features that are not captured by a small sample. Moreover, our flare catalog is relatively small and requiring a perfect balance between classes would lead to only using a small number of non-flaring ARs. A better way is to assign different cost parameters to the two classes. This is the solution we retain here and is a functionality offered by the Support Vector Classification (SVC) class of Scikit-Learn. The cost function of Equation \ref{eq1} is modified the following way:
\begin{equation}
\mathrm{cost} =  \biggl( \frac{1}{2} \mathbf{\omega}^T \mathbf{\omega} + C_1 \sum_{\mathrm{k \in \, positives}} \epsilon_k + C_2 \sum_{\mathrm{k \in \, negatives}} \epsilon_k \biggr)
\end{equation}
where $C_1$ is the cost parameter for the positive class, while $C_2$ is the cost parameter for the negative one.
By setting $C_1 > C_2$, we change the penalty between positive and negative classes and consequently make sure that the classifier does not focus exclusively on the negative one. 

\subsection{Performance metrics}
The results of a binary classifier can be characterized by a confusion matrix, also called a contingency table.
The flaring ARs correctly predicted as flaring are called true positives (TP), the flaring ARs incorrectly predicted as non-flaring are false negatives (FN), the non-flaring ARs correctly predicted as non-flaring are true negatives (TN), and the non-flaring ARs incorrectly predicted as flaring are false positives (FP). From these four quantities, various metrics are computed.

\noindent The precision characterizes the ability of the classifier not to label as positive an example that is negative, and is defined as:
\begin{equation}
\mathrm{precision} = \frac{\mathrm{TP}}{\mathrm{TP}+\mathrm{FP}}
\end{equation}
The recall (also known as the sensitivity or hit rate) characterizes the ability of the classifier to find all of the positive examples:
\begin{equation}
\mathrm{recall} = \frac{\mathrm{TP}}{\mathrm{TP}+\mathrm{FN}}
\end{equation}
Precision and recall are usually anti-correlated: the recall will decrease when the precision increases, and vice-versa.
Therefore, a useful quantity to compute is their harmonic mean, the f1 score:
\begin{equation}
\mathrm{f1}=\frac{2 \times \mathrm{precision} \times \mathrm{recall}}{\mathrm{precision} + \mathrm{recall}}
\end{equation} 
Precision, recall, and f1 score can be computed for both the positive and negative classes: the definitions above are for the positive class, but by replacing TP with TN, and FP with FN (and vice-cersa), we can define similar quantities for the negative class. For instance, the (negative) recall is defined as TN$/$(TN+FP), and is also called specificity or true negative rate.
Another widely used performance metric for a classifier is the accuracy, defined as:
\begin{equation}
\mathrm{accuracy} = \frac{\mathrm{TP}+\mathrm{TN}}{\mathrm{TP}+\mathrm{FN}+\mathrm{TN}+\mathrm{FP}}
\end{equation}
This is the ratio of the number of correct predictions over the total number of predictions.

Many performance metrics are significantly impacted by the class imbalance that skews the distribution of flaring and non-flaring ARs. 
For instance, the accuracy is meaningless: a classifier always predicting that an AR will not flare would result in a very good accuracy even though such a classifier would be quite useless for our purpose.
Similarly, precision and f1-score should not be relied upon. Indeed, by increasing the number of negative examples in the dataset used to test the classifier, while maintaining a constant number of positive examples, we may increase the number of false positives (assuming a constant false positive rate) and therefore artificially lower the precision.
Recall is a more reliable metric as it is insensitive to the number of examples in the majority class.

A better way to estimate the performance of a classifier is to determine how it compares to a given benchmark by computing a skill score.
Skill scores are usually presented in the format of a score value minus the score of a standard forecast divided by a perfect score minus the score of the standard forecast. 
A widely used skill score is the Heidke Skill Score (HSS) \citep[e.g.][]{balch08}. Two different definitions have been applied to the solar-flare prediction field. \citet{bl08} define the HSS as:
\begin{equation}
\mathrm{HSS}_1=\frac{\mathrm{TP}+\mathrm{TN}-\mathrm{N}}{\mathrm{P}}=\mathrm{recall} \times \biggl(2-\frac{1}{\mathrm{precision}}\biggr)
\end{equation}
where P$=$TP$+$FN is the total number of positives, while N$=$TN$+$FP is the total number of negatives.
HSS$_1$ ranges from $-\infty$ to 1 and measures the improvement of the forecast over always predicting that no flare will occur.
A score of 1 means a perfect forecast, and negative scores denote a worse performance than always predicting no flare.
It makes sense to define such a skill score for the prediction of solar flares, as there are many more non-flaring regions than flaring ones and, consequently, beating the ``always non-flaring'' forecast is a stringent requirement.
However, this metric proves unhelpful when comparing the results of different forecasting studies, as it strongly depends on the class-imbalance ratio (N$/$P) of the testing set (we assume a constant ratio for the training set): for instance, if we increase the number of non-flaring active regions N in the testing set while keeping the number of flaring regions P constant, it becomes much more difficult to beat the benchmark because always predicting that no flare will occur gives increasingly better results. Indeed, if N$=1000$ and the classifier has a $98\%$ true negative rate, then TN$=0.98*1000$ and TN$-$N$=-20$. If, now, N$=10000$ and the classifier keeps the same true negative rate, then TN$-$N$=-200$. Because P and TP are kept constant, then HSS$_1$ decreases significantly by virtue of a larger N. This is an undesirable feature for a skill score aimed at inter-group comparison.

On the other hand, \citet{mason10} use a definition of the HSS provided by the Space Weather Prediction Center:
\begin{equation}
\mathrm{HSS}_2 = \frac{\mathrm{TP}+\mathrm{TN}-\mathrm{E}}{\mathrm{P}+\mathrm{N}-\mathrm{E}}
\end{equation}
where E is the expected number of correct forecasts due to chance. In other words, HSS$_2$ measures the improvement of the forecast over a random forecast. E is defined as:
\begin{equation}
\mathrm{E}=\frac{(\mathrm{TP}+\mathrm{FP})\times(\mathrm{TP}+\mathrm{FN})+(\mathrm{FP}+\mathrm{TN})\times(\mathrm{FN}+\mathrm{TN})}{\mathrm{P+N}}
\end{equation}
HSS$_2$ can also be written as:
\begin{equation}
\mathrm{HSS}_2 = \frac{2 \times [(\mathrm{TP}\times \mathrm{TN})-(\mathrm{FN}\times \mathrm{FP})]}{\mathrm{P} \times(\mathrm{FN}+ \mathrm{TN})+(\mathrm{TP}+ \mathrm{FP})\times \mathrm{N}}
\end{equation}

\noindent Another skill score used by \citet{mason10} is the Gilbert skill score GS:
\begin{equation}
\mathrm{GS} = \frac{\mathrm{TP}-\mathrm{CH}}{\mathrm{TP}+\mathrm{FP}+\mathrm{FN}-\mathrm{CH}}
\end{equation}
where CH is the number of TP obtained by chance. CH is defined as:
\begin{equation}
\mathrm{CH}=\frac{(\mathrm{TP}+\mathrm{FP})\times(\mathrm{TP}+\mathrm{FN})}{\mathrm{P+N}}
\end{equation}
Therefore CH only includes true positives, while E also includes true negatives. 

\begin{figure}
\epsscale{1.0}
\plotone{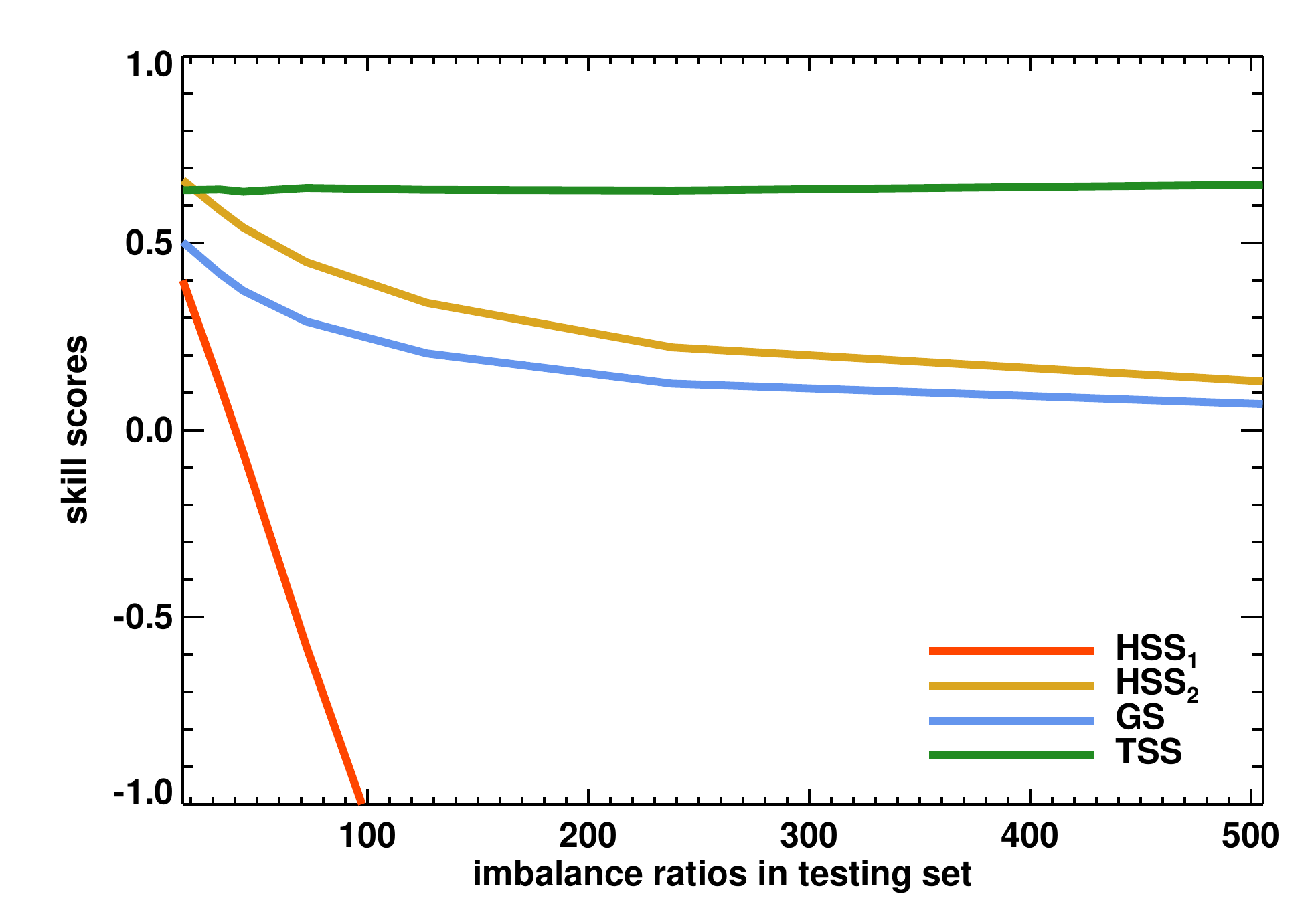}
\caption{Dependence of the HSS$_1$, HSS$_2$, GS, and TSS, on the class-imbalance ratio N/P in the testing set, for a fixed ratio in the training set of $16.5$ and the SVM parameters listed in Section \ref{section:results}. The dataset was built for the operational case and contains the first 16 features listed in Table \ref{tbl-3}.\label{fig1}}
\end{figure}

Similarly to HSS$_1$, HSS$_2$ and GS exhibit some dependence on the class-imbalance ratio of the testing set. To alleviate this problem, \citet{bloomfield12} suggest the use of the True Skill Statistic (TSS), sometimes called the  Hansen-Kuipers skill score or Peirce skill score \citep{woodcock76}, defined as:
\begin{equation}
\mathrm{TSS}=\frac{\mathrm{TP} \times \mathrm{TN}-\mathrm{FP} \times \mathrm{FN}}{\mathrm{P} \times \mathrm{N}}=\frac{\mathrm{TP}}{\mathrm{TP}+\mathrm{FN}}-\frac{\mathrm{FP}}{\mathrm{FP}+\mathrm{TN}} \label{TSS}
\end{equation} 
The TSS is the difference between the recall and false alarm rate, \mbox{i.e.} the recall plus the specificity minus one. This is equivalent to one minus the false alarm rate minus the false negative rate (defined as FN/(FN+TP)): therefore, the best TSS is 1 and any misclassification, both of positive or negative examples, reduce this score accordingly. Equation \ref{TSS} does not follow the standard skill score format, but the TSS can be written in a similar fashion. The TSS ranges from $-1$ to $+1$ and has the desirable characteristic of equitability \citep{richardson00}: random or constant forecasts score zero, perfect forecasts score 1, and forecasts that are always wrong score $-1$. Another desirable feature of the TSS is that it is unbiased with respect to the class-imbalance ratio (see Figure \ref{fig1}), meaning that it does not change for unequal trials. Following \citet{bloomfield12} we believe that the TSS should be the metric of choice when comparing the performances of various classifiers for solar flare forecast. \citet{manzato05} reached a similar conclusion with respect to weather forecasts of thunderstorms (rare events), and the author demonstrates that the TSS is a good measure of the overall classifier quality. 

Even though we calculate all of the metrics previously described, we believe that the TSS is the most meaningful for comparison purposes. Ideally, comparing the performances of classifiers from various groups would require using the same class-imbalance ratio in all of the testing sets, which has not been done in the articles surveyed.

A potential issue with the TSS is that it treats false positives FP (false alarms) and false negatives FN in the same way, irrespective of the difference in consequences: not predicting a flare that occurs (false negative) may be more costly than predicting a flare that does not occur (false alarm). Indeed, in the case of a satellite partly shielded to withstand an increase in energetic particles following a solar flare, the cost of a false alarm is the price paid to rotate the satellite so that the shielded part faces the particle flow, while the cost of a false negative may be the breakdown of the satellite. Both costs are asymmetric.

Figure \ref{fig1} highlights how the different skill scores presented in this section vary with the class imbalance ratio N/P: the ratio is set to $16.5$ in the training set, and varies in the testing set. The other SVM parameters are set to the values listed in the next section. Only the TSS does not depend on the testing-set imbalance ratio, while both HSS$_2$ and GS converge toward zero for large N/P ratios. HSS$_1$ becomes more and more negative with N/P and exhibits the strongest dependence.

\section{Results}
\label{section:results}

We use the following parameters to train the SVM: $C=4$, $\gamma=0.075$, tolerance of $10^{-8}$ on the relative change in the cost function, class-imbalance ratio N/P of $16.5$ for both the training and testing sets, and a ratio between $C_1$ (penalty parameter for the positive class) and $C_2$ (penalty parameter for the negative class) of $2.0$ (for the results presented in Table \ref{tbl-1} and Figures \ref{fig1} and \ref{fig4}) and $15.0$ (for the results presented in Table \ref{tbl-2}). The catalog of examples is randomly divided into a testing set ($30\%$ of the number of examples) and a training set ($70\%$). We maintain N/P close to the ratio in the testing set of \citet{ahmed} to make comparison with their results meaningful (with respect to the precision, the f1 score, HSS$_1$, HSS$_2$, and GS).
It is important to note that our results are tailored for comparison with \citet{ahmed} (despite the fact that they include C-class flares in their dataset) and that, even though we provide comparisons with the results of other groups, performance metrics other than the TSS and the recall are not really meaningful as previously explained: the N/P ratio varies too much across the literature. For instance, \citet{mason10} have a ratio of $260.5$, while \citet{bl08} have a ratio of $9.9$.

The first step in training the SVM is to pre-process the training examples in our dataset by discarding SHARP entries for which the QUALITY keyword is greater than 65536 (or 10000 in hexadecimal), corresponding to unreliable Stokes vectors (observables that were produced in bad conditions). Then, we scale our features: to improve the performances of the feature selection algorithms and the classifier, it is required to have features that lie within similar ranges. Therefore, we subtract from every feature its median value and divide by its standard deviation. The performance of the SVM seems to be somewhat impacted by the specifics of feature scaling. Subtracting the median rather than the mean appears to slightly improve the results, which is likely due to the lesser sensitivity of the median value to outliers.

\subsection{Feature selection}

\begin{figure}
\epsscale{1.0}
\plotone{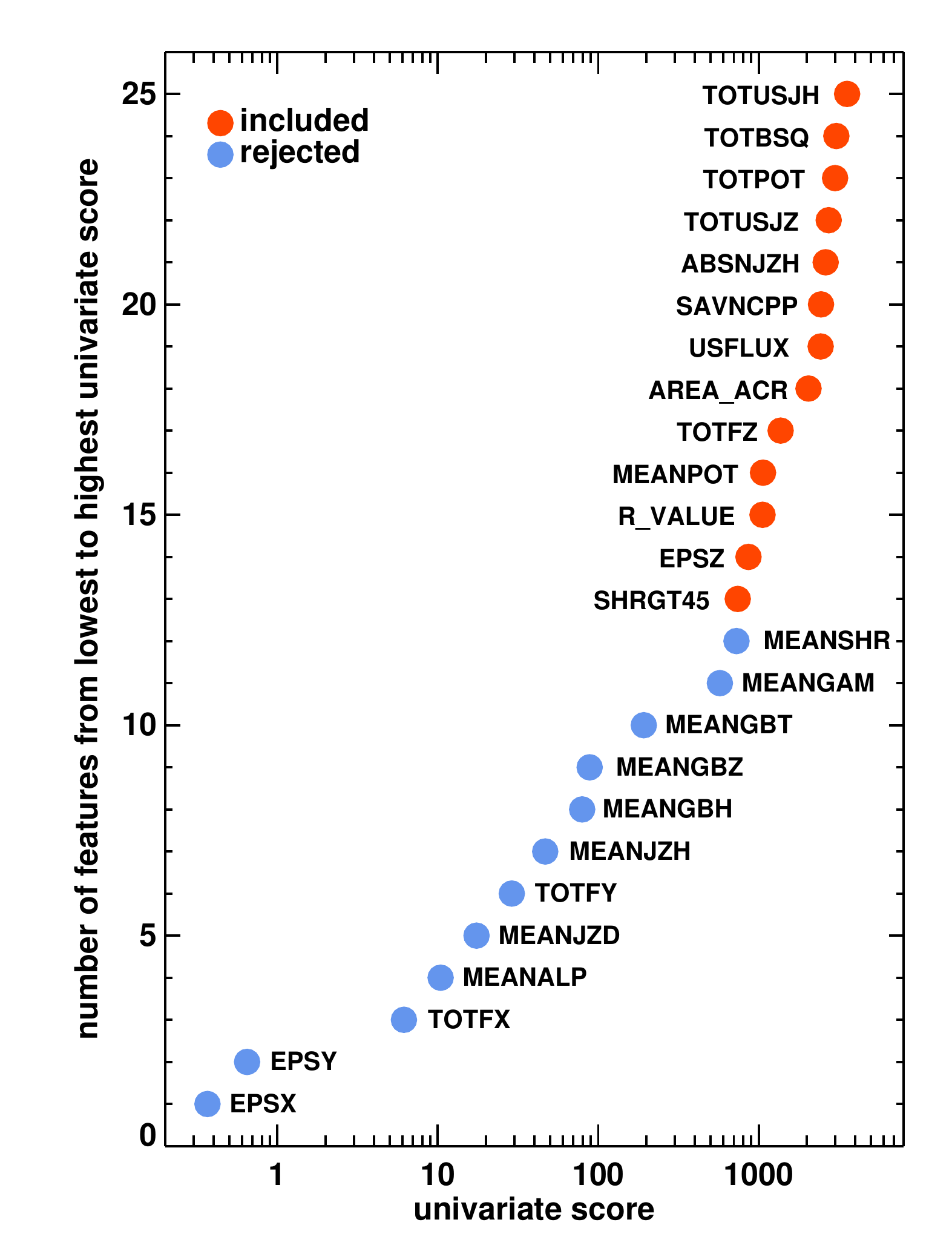}
\caption{Univariate Fisher score for the 25 features of the SHARP parameters in the operational case. The features included in our forecast algorithm are in orange. \label{fig3}}
\end{figure}

Not all of the features are useful predictors of flaring activity. The F-scores $F(i)$ clearly show that some features are, in fact, quite useless. Figure \ref{fig3} displays the scores obtained for the operational case (\mbox{i.e.} the negative class is for non-flaring ARs in the 24 hours following the sample time), and after the features have been scaled. 
The feature with the highest F-score is the total unsigned current helicity. The absolute value of the net current helicity is also highly ranked. Many studies \citep[e.g.][]{georg09} suggest the build-up of magnetic helicity is intimately linked with flare productivity. Magnetic fluxes, both the total unsigned flux and the flux summed along the high-gradient active region polarity inversion line, are also high scoring parameters. \citet{schrijver07} found that large flares are always associated with high-gradient active region polarity inversion lines. We confirm this finding. Our sample shows a similar linear correlation betwen R and total unsigned flux as described in Figure 2 of \citet{schrijver07}. Finally, Table \ref{tbl-3} and Figure \ref{fig3} clearly show that the most valuable parameters in terms of flare-prediction are those that calculate sums, rather than means, of various physical quantities. \citet{welsch09} also find a similar result.  

\begin{figure}
\epsscale{1.0}
\plotone{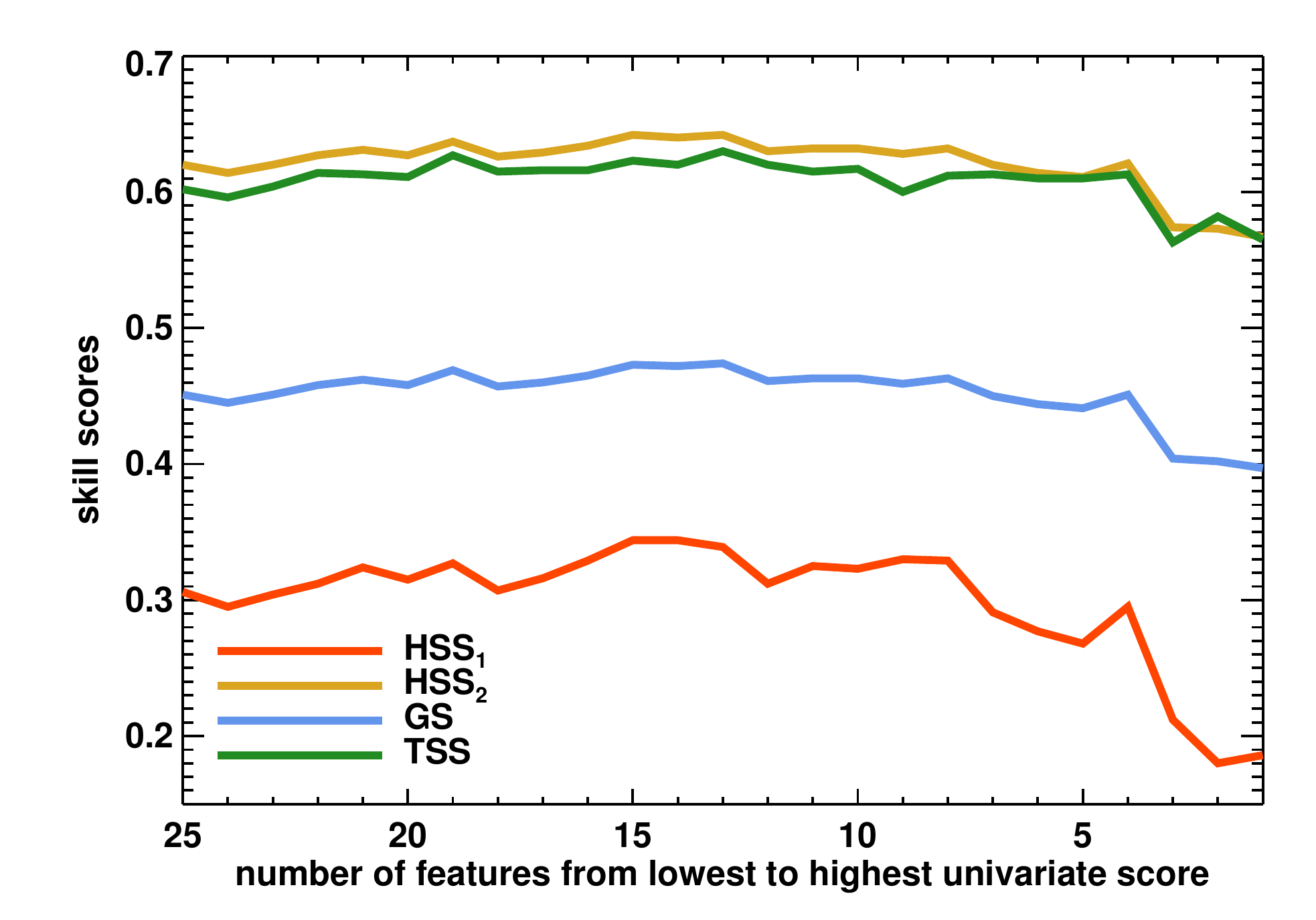}
\caption{Skill scores as a function of the number of features that characterize the active regions in the operational case and with the SVM parameters listed in Section \ref{section:results}.\label{fig4}}
\end{figure}

Figure \ref{fig4} highlights how the skill scores of our SVM classifier vary with the number of features. To create this plot, we start with 25 features. Then we remove one feature, that with the lowest univariate score, at a time. In the end, only the total unsigned current helicity remains. The rms variation for the TSS on Figure \ref{fig4}, obtained from 1000 iterations of the training and testing of the SVM, is about $0.048$. Consequently, the small changes in the TSS for a feature number between 4 and 25 are well within the 1-sigma error bar. However, we wish to select a subset of features to use with our forecasting algorithm (\mbox{e.g.}, to speed up computations by discarding irrelevant features). Also, we fit the TSS with a third-order polynomial and find that it peaks at 13 features, which are the first 13 listed in Table \ref{tbl-3}. Again, all of these features, except for AREA\_ACR and R, can only be derived from vector magnetic field data. The results in the rest of this paper are obtained with this specific subset of 13 parameters.

Univariate selection does not account for correlation between the features, and the way they may potentially interact. For instance, two features may be poor flare predictors when considered separately, but may be quite useful when combined. Correlation between features can be an issue for some ML algorithms. The SVM algorithm, however, is not sensitive to feature correlation. Some of the 13 features we retain for the SVM are in fact highly correlated, which is not surprising as they are all good flare predictors. For instance, the Pearson linear correlation coefficient between the total unsigned current helicity and the total magnitude of the Lorentz force (the two highest-ranked features) reaches $97.5\%$, while this coefficient is only $-1.7\%$ between the total unsigned current helicity and the sum of the x-component of the normalized Lorentz force (the highest and lowest-ranked features). \citet{guyon03} point out that high feature correlation does not mean absence of feature complementarity.

It is noteworthy that using only the 4 highest-ranking parameters --- the total unsigned current helicity, total magnitude of the Lorentz force, total photospheric magnetic free energy density, and total unsigned flux --- gives roughly the same TSS score as the top 13 combined. \citet{lb07} also conclude that increasing the number of parameters beyond a few adds little value. Their five top-performing parameters are: the total photospheric excess magnetic energy, total unsigned vertical current, total unsigned vertical heterogeneity current, total unsigned current helicity, and total unsigned flux. In our study, these five features all have a high F-score. Further, both our study and \citet{lb07} find the total photospheric magnetic free energy, total unsigned vertical current, and total unsigned current helicity to be important parameters for flare prediction. 

\subsection{Forecasting algorithm performance}

As previously mentioned, the results in this section are obtained with 13 features. For comparison with \citet{ahmed}, we combine their Tables 6 (which lists their best results) and 4 to recreate their contingency tables (from Table 4 we get P and N for the testing sets, and from TPR, FPR, TNR, and FNR in their Table 6 we derive the performance metrics). For \citet{mason10} we use the contingency table they provide in Table 2. The HSS$_2$ value we derive is different from the one quoted in their Conclusion section, but it matches the one calculated by \citet{bloomfield12} also based on Table 2 of \citet{mason10}. For \citet{yu09} we use the results provided by \citet{bloomfield12} in their Table 4, which come from a private communication. For \citet{bloomfield12} we use their Table 4 and only the entries for flares larger than M1.0 class. \citet{song09} listed their results in Table 8, but they only have 55 examples in their testing set (for the M-class flare prediction), which means that their results most likely have large error bars (not provided).

The SVM is randomly initialized, and the separation between training and testing examples is also random (although the ratio of 70\% training examples to 30\% testing examples is always maintained). We repeat the training and testing phases 1000 times to provide the means and standard deviations in Tables \ref{tbl-1} and \ref{tbl-2}.

\begin{deluxetable}{cccccccccc}
\tabletypesize{\scriptsize}
\tablecaption{Flare prediction capabilities with 13 features compared to other studies, and with the SVM parameters selected to achieve the highest HSS$_2$. \label{tbl-1}}
\tablewidth{0pt}
\tablehead{
\colhead{Metric} & \colhead{Segmented} &  \colhead{Operational} & \colhead{Mason} & \colhead{Ahmed} & \colhead{Ahmed} & \colhead{Barnes} & \colhead{Bloomfield} & \colhead{Yu} & \colhead{Song}
}
\startdata
Time interval (no flare) & 48h             & 24h             & 6h       & 48h   & 24h   & 24h   & 24h   & 48h  & 24h\\
class-imbalance ratio    & 16.5            & 16.5            & 260      & 15.85 & 16.58 & 9.92  & 26.5  & NA   & 2.23\\
Accuracy                 & 0.973$\pm$0.003 & 0.962$\pm$0.004 & 0.694    & 0.975 & 0.963 & 0.922 & 0.830 & 0.825& 0.873\\
Precision (positive)     & 0.797$\pm$0.050 & 0.690$\pm$0.049 & 0.008    & 0.877 & 0.740 & NA    & 0.146 & 0.831& 0.917\\
Precision (negative)     & 0.983$\pm$0.003 & 0.978$\pm$0.003 & 0.998    & 0.980 & 0.972 & NA    & NA    & NA   & 0.860\\
Recall (positive)        & 0.714$\pm$0.048 & 0.627$\pm$0.049 & 0.617    & 0.677 & 0.523 & NA    & 0.704 & 0.817& 0.647\\
Recall (negative)        & 0.989$\pm$0.003 & 0.983$\pm$0.004 & 0.695    & 0.994 & 0.989 & NA    & NA    & NA   & 0.974\\
f1 (positive)            & 0.751$\pm$0.032 & 0.656$\pm$0.035 & 0.015    & 0.764 & 0.613 & NA    & 0.242 & NA   & 0.758\\
f1 (negative)            & 0.986$\pm$0.002 & 0.980$\pm$0.002 & 0.819    & 0.987 & 0.989 & NA    & NA    & NA   & 0.913\\
HSS$_1$                  & 0.528$\pm$0.062 & 0.342$\pm$0.071 & -78.9    & 0.581 & 0.339 & 0.153 & NA    & NA   & 0.588\\
HSS$_2$                  & 0.737$\pm$0.034 & 0.636$\pm$0.037 & 0.008    & 0.751 & 0.594 & NA    & 0.190 & 0.650& 0.676\\
Gilbert skill score      & 0.585$\pm$0.043 & 0.467$\pm$0.039 & 0.004    & 0.601 & 0.422 & NA    & NA    & NA   & 0.510\\
\hline													       
TSS                      & 0.703$\pm$0.047 & 0.610$\pm$0.048 & 0.312    & 0.671 & 0.512 & NA    & 0.539 & 0.650& 0.620\\
\enddata
\end{deluxetable}

The comparison of our results with other papers in Table \ref{tbl-1} should be taken with a grain of salt, as different groups define positive and negative classes differently, cover different time intervals, have a different N/P ratio in their training and testing sets, cover a different range of solar longitudes and latitudes, and include (or not) C-class flares. Therefore this table should by no means be treated as an exact comparison. Another caveat is that while we provide standard deviations for all of our results, other groups only provide their average or best values. \citet{ahmed} do provide the standard deviations for their HSS$_2$ skill scores ($0.01$ and $0.02$ depending on the table). These values are smaller than ours because of the small number of examples we have compared to them: they include 315561 non-flaring examples, and 16864 flaring ones in their operational form. The small number of flares in our catalog also means that there is an additional risk for our results not to generalize well.

\begin{figure}
\epsscale{1.0}
\plotone{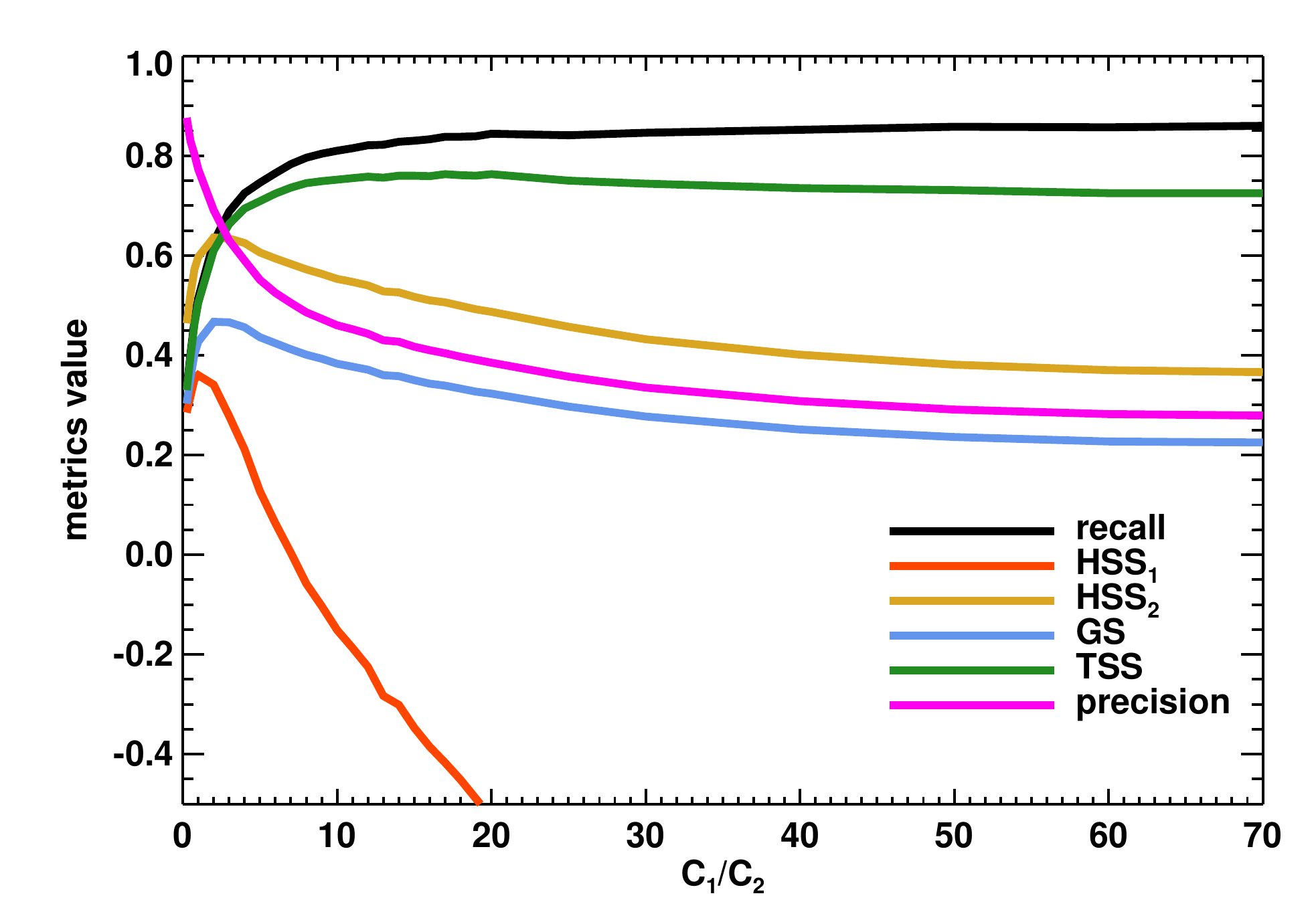}
\caption{Performance metrics as a function of the ratio of the soft parameters for the positive and negative classes, $C_1/C_2$, in the operational case, and for the SVM parameters listed in Section 5.\label{fig5}}
\end{figure}

Table \ref{tbl-1} lists our results when the SVM parameters are selected in order to maximize HSS$_2$; Table \ref{tbl-2} lists our results when the SVM parameters are selected in order to maximize the TSS. The main difference is a change in the $C_1/C_2$ ratio.  We use a relatively high $C_1/C_2$ ratio to maximize the TSS and reduce the misclassification of positive examples. Figure \ref{fig5} shows how this ratio impacts the performance metrics (all other parameters being held constant): the peak values of the different metrics are not located at the same $C_1/C_2$ ratios. As previously mentioned, the TSS should be the preferred metric when comparing the skills of various classifiers, and therefore we believe Table \ref{tbl-2} is more representative of our classifier's performance. We again stress that false negatives FN in solar flare prediction can be quite costly to astronauts and satellites orbiting the Earth, and that therefore it seems preferable to accept a bias in the classifier: it is better to aim for a large recall (thus minimizing the false negative rate, equal to one minus the recall), even if this results in a poor precision. A large TSS goes hand-in-hand with a large recall.

From Tables \ref{tbl-1} and \ref{tbl-2}, it appears that our forecasting algorithm exhibits better performances than the others listed, when comparing the TSS. Compared to \citet{ahmed}, our classifier performs better both in operational and segmented modes. To be fair, \citet{ahmed} did not try to maximize the TSS (which they did not use at all), and instead focused on HSS$_2$ and other metrics: this, at least partly, explains their lower TSS. Our higher score could also result from a better set of features derived from vector magnetograms. It is difficult to draw firm conclusions from this comparison, as we use a different forecasting algorithm, a different database, and we only consider major flares while they also forecast C-class ones. The highest TSS we reach in operational mode is $0.76$, while it is $0.82$ in segmented mode. These numbers are quite good, but result in a low precision (only $50\%$ in segmented and $42\%$ in operational modes). Therefore, with the set of SVM parameters designed to maximize the TSS, our forecasting algorithm overclassifies flaring active regions, producing a lot of false positives. Finally, it is noticeable that \citet{song09} also obtain a high TSS, but their testing set only includes 55 examples, which means that their results probably come with large error bars and may not generalize very well to larger datasets.

\begin{deluxetable}{ccc}
\tabletypesize{\scriptsize}
\tablecaption{Flare prediction capabilities with 13 features, and with the SVM parameters selected to achieve the highest TSS. \label{tbl-2}}
\tablewidth{0pt}
\tablehead{
\colhead{Metric} & \colhead{Segmented} &  \colhead{Operational} }
\startdata
Time interval (no flare) & 48h             & 24h             \\
class-imbalance ratio    & 16.5            & 16.5            \\
Accuracy                 & 0.943$\pm$0.006 & 0.924$\pm$0.007 \\
Precision (positive)     & 0.501$\pm$0.041 & 0.417$\pm$0.037 \\
Precision (negative)     & 0.992$\pm$0.002 & 0.989$\pm$0.003 \\
Recall (positive)        & 0.869$\pm$0.036 & 0.832$\pm$0.042 \\
Recall (negative)        & 0.947$\pm$0.007 & 0.929$\pm$0.008 \\
f1 (positive)            & 0.634$\pm$0.033 & 0.554$\pm$0.033 \\
f1 (negative)            & 0.969$\pm$0.003 & 0.958$\pm$0.004 \\
HSS$_1$                  & -0.008$\pm$0.142 & -0.348$\pm$0.183 \\
HSS$_2$                  & 0.606$\pm$0.035 & 0.517$\pm$0.035 \\
Gilbert skill score      & 0.436$\pm$0.036 & 0.350$\pm$0.032 \\
\hline							     
TSS                      & 0.817$\pm$0.034 & 0.761$\pm$0.039 \\
\enddata
\end{deluxetable}

\section{Conclusion}

In the absence of a definitive physical theory explaining the flaring mechanism of an active region, the best hope for forecasting solar flares lies in finding an empirical relationship between some well chosen features of active regions and flare productivity. To find such a relationship, we use 25 active-region parameters, calculated by sampling from a database of 1.5 million active region patches of vector magnetic field data, and a support vector machine algorithm. We only aim at forecasting major solar flares (M1.0-class and above). Most forecasting efforts so far have been based on line-of-sight magnetograms, with the notable exception of \citet{lb07} who used ground-based vector magnetograms. However, the use of HMI vector magnetograms provides a much larger database and a more uniform quality allowed by space-based observations. We also access a number of active-region features (25) larger and more diverse than most groups.

Solar flare prediction is a strongly imbalanced problem, as there are many more non-flaring examples than flaring ones. Based on \citet{bloomfield12}, who highlight the dependence of several skill scores on this class imbalance, we favor the true skill statistic, TSS, to compare our forecasting algorithm with others' results. The TSS has several desirable features in a performance metric. A high TSS implies a low false negative rate, which is very relevant to solar-flare forecasting as false negatives can prove quite costly in terms of the well-being of astronauts or satellites orbiting the Earth.

Here, we obtain high TSS values for the operational and segmented modes: the positive class (flaring active region) and negative one (non-flaring active region) are defined following \citet{ahmed}, and we use the same class-imbalance ratio as them to make any comparison with their results meaningful. The error bars on our skill scores are larger than in their paper because we have fewer flaring active regions, an unfortunate consequence of the quietness of solar cycle 24. Moreover, we do not include C-class flares in our dataset. 

With these caveats in head, the TSS we obtain are larger than in the papers studied. This is partly due to the fact that other groups focus on reaching high scores for performance metrics other than the TSS. Maximizing some scores may result in lower values for other metrics. Another explanation for our higher TSS may be the use of vector magnetograms: they give access to the field topology and allow us to compute active-region features that cannot be measured based on line-of-sight magnetic-field maps. This is the first time a large dataset of vector magnetograms have been used to forecast solar flares.

Despite the relative success of our forecasting algorithm, its predictive capabilities are far from perfect. For instance, the false negative rate is $28.6\%$ in the segmented mode with the parameters used to produce Table \ref{tbl-1}, which means that a significant number of flaring active regions are still forecasted as non-flaring. We do manage to obtain a higher TSS, listed in Table \ref{tbl-2}, with values as high as $0.82$ in segmented mode, resulting in a lower false negative rate at $13\%$, but the other performance metrics are also low (\mbox{e.g.}, the precision is only $50\%$). The false alarm rate (false positive rate) is always artificially low due to the large number of negative examples in our catalog.

Although we calculate a diverse set of 25 parameters, we find that using only the 4 parameters with the highest F-score --- the total unsigned current helicity, total magnitude of the Lorentz force, total photospheric magnetic free energy density, and total unsigned vertical current --- gives roughly the same TSS score as the top 13 combined. Both our study and \citet{lb07} find the magnetic energy, vertical current, and current helicity are useful physical quantities for flare prediction. It is also noteworthy that these most valuable parameters calculate sums, rather than means, of various physical quantities. \citet{welsch09}, who studied the relationship between photospheric flows and flaring activity, also found that extensive parameters (i.e. those that increase with system size) are more correlated with flaring behavior than intensive parameters (i.e. those that do not increase with system size). 

Others have also found that a few select parameters contain most of the information in the entire parameter set. \citet{ahmed} found that their 6 most valuable parameters produced a prediction capability comparable to their entire set of 21 parameters. \citet{lb07} also conclude that increasing the number of parameters beyond a few adds little value. This suggests that the photospheric magnetic field contains limited information and various permutations of this information are not useful. As it has been emphasized in the past \citep[e.g.][]{lb03,lb07}, it is not clear that using photospheric magnetic field data alone will allow us to significantly improve these forecasting capabilities.

In the near future, we shall work on several topics to try to improve the forecasting performances: we plan to combine the SVM with other algorithms, like the $k$-nearest neighbors \citep[following][]{li07} and we plan to use multi-class classifiers to not only forecast a flare occurrence, but also to attempt to predict whether it will be an M or X-class flare. The active region parameters are fairly sensitive to which pixels contribute to their calculation. Therefore we would also like to test different masks to yield the strongest pre-flare signature per AR parameter.

\acknowledgments
This work was supported by NASA Grant NAS5-02139 (HMI). The data used here
are courtesy of NASA/SDO and the HMI science team, as well as the Geostationary Satellite System (GOES) team.


\end{document}